\documentclass[useAMS,usenatbib,usegraphicx]{mn2e}
\usepackage{times}
\usepackage{amsmath}
\usepackage{amssymb}

\newcommand*{\cii}{\text{[C\,\textsc{ii}]}}

\newcommand*{\herschel}{\emph{Herschel}}
\newcommand*{\J}[2]{\ensuremath{\text{\emph{J}}\! =\! #1\! \rightarrow\! #2}}
\newcommand*{\K}{\textsc{k}}
\newcommand*{\kms}{\text{km\,s\(^{-1}\)}}
\newcommand*{\lir}{\ensuremath{L_\text{IR}}}

\newcommand*{\lsun}{\ensuremath{\text{L}_{\sun}}}
\newcommand*{\msun}{\ensuremath{\text{M}_{\sun}}}

\newcommand*{\R}[4]{\ensuremath{r_{#1 \text{--} #2 / #3 \text{--} #4}}}
\newcommand*{\Tb}{\ensuremath{T_\text{b}}}
\newcommand*{\Td}{\ensuremath{T_\text{d}}}
\newcommand*{\um}{\ensuremath{\umu \text{m}}}
\newcommand*{\uJy}{\ensuremath{\umu \text{Jy}}}
\newcommand*{\mJy}{\text{mJy}}
\newcommand*{\xunits}{
  \ensuremath{\msun \left(\K{} \, \kms \text{pc}^2 \right)^{-1}}
}

\def\gs{\mathrel{\raise0.35ex\hbox{$\scriptstyle >$}\kern-0.6em
\lower0.40ex\hbox{{$\scriptstyle \sim$}}}}
\def\ls{\mathrel{\raise0.35ex\hbox{$\scriptstyle <$}\kern-0.6em
\lower0.40ex\hbox{{$\scriptstyle \sim$}}}}
\def\m@th{\mathsurround=0pt }
\def\eqalign#1{\null\,\vcenter{\openup1\jot \m@th
 \ialign{\strut\hfil$\displaystyle{##}$&$\displaystyle{{}##}$\hfil
 \crcr#1\crcr}}\,}

\title[Far-infrared spectroscopy of a lensed starburst]{
Far-infrared spectroscopy of a lensed starburst: a blind redshift
from \herschel
}
\author[George et al.]{R.\,D.~George,\(^{\! 1}\)
R.\,J.~Ivison,\(^{\! 2,1}\)
R.~Hopwood,\(^{\! 3,4}\)
D.\,A.~Riechers,\(^{\! 5}\)
R.\,S.~Bussmann,\(^{\! 5}\)
P.~Cox,\(^{\! 6}\) \and
S.~Dye,\(^{\! 7}\)
M.~Krips,\(^{\! 6}\)
M.~Negrello,\(^{\! 8}\)
R.~Neri,\(^{\! 6}\)
S.~Serjeant,\(^{\! 4}\)
I.~Valtchanov,\(^{\! 9}\)
M.~Baes,\(^{\! 10}\) \and
N.~Bourne,\(^{\! 7}\)
D.\,L.~Clements,\(^{\! 3}\)
G.~De~Zotti,\(^{\! 8,11}\)
L.~Dunne,\(^{\! 12}\)
S.\,A.~Eales,\(^{\! 13}\)
E.~Ibar,\(^{\! 14}\) \and
S.~Maddox,\(^{\! 12}\)
M.\,W.\,L.~Smith,\(^{\! 13}\)
E.~Valiante\(^{13}\)
and
P.~van~der~Werf\(^{15}\)
\vspace*{1mm}\\
\(^1\) Institute for Astronomy, University of Edinburgh,
         Blackford Hill, Edinburgh EH9 3HJ\\
\(^2\) UK Astronomy Technology Centre,
         Royal Observatory, Blackford Hill, Edinburgh EH9 3HJ\\
\(^3\) Physics Dept, Imperial College London, South
         Kensington Campus, London SW7 2AZ\\
\(^4\) Dept of Physical Sciences, The Open University, Milton Keynes\\
\(^5\) Dept of Astronomy, Space Science Building,
         Cornell University, Ithaca NY\,14853-6801, USA\\
\(^6\) Institut de Radioastronomie Millim\'etrique,
         300 rue de la Piscine, 38406 Saint-Martin d'H\`eres, France\\
\(^7\) School of Physics \& Astronomy, University of
         Nottingham, University Park, Nottingham NG7 2RD\\
\(^8\) INAF, Osservatorio Astronomico di Padova, I-35122 Padova, Italy\\
\(^{9}\) Herschel Science Centre, European Space Astronomy Centre, ESA,
            28691 Villanueva de la Ca\~nada, Spain\\
\(^{10}\) Sterrenkundig Observatorium, Universiteit Gent, Krijgslaan
            281 S9,  B-9000 Gent, Belgium\\
\(^{11}\) SISSA, via Bonomea 265, I 34136 Trieste, Italy\\
\(^{12}\) Dept of Physics and Astronomy, University of Canterbury,
            Private Bag 4800, Christchurch, New Zealand\\
\(^{13}\) School of Physics \& Astronomy, Cardiff University,
            Queen's Buildings, The Parade 5, Cardiff CF24 3AA\\
\(^{14}\) Pontificia Universidad Cat\'olica de Chile,
            Departamento de Astronom\'ia y Astrof\'isica,
            Vicu\~na Mackenna 4860, Casilla 306, Santiago 22, Chile\\
\(^{15}\) Leiden Observatory, Leiden University,
            P.O. Box 9513, 2300 RA Leiden, Netherlands}
\date{
Accepted 2013 Xxxxxx X. Received 2013 Xxxxxx X; in original form 2013
Xxxxxx X
}
\pagerange{000--000}

\begin{document}

\maketitle

\begin{abstract}
We report the redshift of HATLAS\,J132427.0+284452 (hereafter
HATLAS\,J132427), a gravitationally lensed starburst galaxy, the
first determined `blind' by the \emph{Herschel Space Observatory}.
This is achieved via the detection of \cii{} consistent with
\(z=1.68\) in a far-infrared spectrum taken with the SPIRE Fourier
Transform Spectrometer.  We demonstrate that the \cii{} redshift is
secure via detections of CO \J{2}{1} and \(3\!\rightarrow\!2\) using
the Combined Array for Research in Millimeter-wave Astronomy and the
Institut de Radioastronomie Millim\'{e}trique's Plateau de Bure
Interferometer.  The intrinsic properties appear typical of
high-redshift starbursts despite the high lensing-amplified fluxes,
proving the ability of the FTS to probe this population with the aid
of lensing.  The blind detection of \cii{} demonstrates the potential
of the SAFARI imaging spectrometer, proposed for the much more
sensitive SPICA mission, to determine redshifts of multiple dusty
galaxies simultaneously without the benefit of lensing.
\end{abstract}


\section{Introduction}
\label{introduction}
Through surveys using the Submillimetre Common-User Bolometer Array
\citep[SCUBA --][]{1999MNRAS.303..659H} came the discovery of a
population of dust-obscured, submillimetre- (submm-)bright galaxies
\citep{2002PhR...369..111B}.  Analogous to ultraluminous infrared
galaxies (ULIRGs) in the local neighbourhood, these distant, gas-rich,
intensely star-forming galaxies emit the bulk of their radiation in
the rest-frame far-infrared (far-IR) waveband.

Surveys with the latest generation of far-IR- and submm-wavelength
facilities, in particular the 3.5\,m \emph{Herschel Space Observatory},
have enabled us to image many more of these dusty star-forming
galaxies (DSFGs) than was previously possible, over vastly larger
areas, in up to five filters simultaneously, allowing us to probe rare
populations that were not well-sampled in previous far-IR surveys.

The discovery of a bright, lensed population of DSFGs, intrinsically
below the \herschel{} detection limit
\citep[e.g.][]{2010Sci...330..800N}, has confirmed that far-IR imaging
is an efficient method by which to select strongly lensed objects and
due to the sometimes high magnification \citep[e.g.
\(\mu = 37.5 \pm 4.5\);][]{2011ApJ...742...11S},
has permitted detailed studies of individual galaxies  \citep[e.g.][]
{2011ApJ...740...63C, 2013ApJ...772..137I}.

High-resolution studies of individual DSFGs typically rely upon
interferometric CO-line observations, necessitating precise knowledge
of their redshifts.  These remain notoriously difficult to obtain, due
to the difficulty in pinpointing their positions, and their faintness
in the optical.  However, the lensed DSFG population has proved to be
within reach of a range of new ground-based, broadband spectroscopic
instrumentation \citep[e.g.][]{2012ApJ...752..152H}; very recently,
\citet{2013ApJ...767...88W} have demonstrated the power of ALMA,
particularly at $z>4$ where the 3-mm atmospheric window always
contains a line.

Here, we present the first blind determination of a redshift using
far-IR spectroscopic observations from space -- a key step towards
demonstrating the feasibility of integral-field far-IR spectroscopy as
planned for the joint JAXA-ESA mission, SPICA
\citep[e.g.][]{2009ExA....23..193S}.  Throughout the paper we use
WMAP7 cosmology \citep{2011ApJS..192...18K} with
\(H_0 = 70.4 \, \kms \, \text{Mpc}^{-1}\), \(\Omega_\text{m} = 0.272\)
and \(\Omega_\Lambda = 0.728\).

\section{Discovery observations}
\label{basic}

The North Galactic Pole (NGP) is one of a number of fields observed as
part of the \emph{Herschel}-ATLAS \citep{2010PASP..122..499E} with
\herschel{} \citep{2010A&A...518L...1P}.  The acquisition and reduction
of parallel-mode data from SPIRE \citep{2010A&A...518L...3G} and PACS
\citep{2010A&A...518L...2P} are described in detail by
\citet{2010MNRAS.409...38I}, \citet{2011MNRAS.415..911P} and
\citet{2011MNRAS.415.2336R}.

The SPIRE imaging led quickly to the selection of
HATLAS\,J132427.0+284452 as a bright \(\left(S_{350\um} \gtrsim
200 \, \text{mJy;}\right.\) all bands \(\left. \geqslant\,30\,
\sigma\right)\), potentially distant \(\left(\left\{S_{250\um},
S_{500\um}\right\} < S_{350\um}\right)\), lensed starburst.  This
source is included in a number of follow-up campaigns, where we have
been studying the brightest of these objects
\citep[e.g.][]{2010Sci...330..800N, 2012ApJ...753..134F,
2013A&A...551A.115O}.

The surface density of \(S_{350\um} \gtrsim 300 \, \mJy\) lens
candidates is likely around 1/40\,deg\(^{-2}\) (M.~Negrello, priv.\
comm.), plausibly placing this source among the most strongly lensed
known DSFGs.

\section{Detailed follow-up observations}
\label{followup}

Much deeper PACS data were subsequently obtained from the \textsc{ot1}
programme, \textsc{ot1\_rivison\_1}, recording data simultaneously at
100 and 160\,\um\ for a total of 6\,min, and reaching \(\sigma \approx
10\) and 12\,mJy respectively.

Approximately 30\,min of high-resolution 7\,GHz continuum data were
acquired using the National Radio Astronomy Observatory's Jansky Very
Large Array (NRAO's VLA) during 2011 June.  The observation was
performed in A configuration, with $64\times 2$-MHz channels in each
of two intermediate frequencies (IFs), each IF with dual polarisation,
recording data every 1\,s.  3C\,286 was observed every few minutes to
determine accurate complex gain solutions and bandpass corrections,
and to set the absolute flux density scale.  Using natural weighting,
the resulting map has a circular 0.3-arcsec synthesised beam (full
width at half maximum, FWHM) and an r.m.s.\ noise level of
\(10 \, \uJy \, \text{beam}^{-1}\).

870-\um\ imaging data were obtained using the Submillimeter
Array (SMA) as part of programme 2011B-S044, with 9.7\,hr of total
integration time in the compact, extended, and very extended array
configurations, with baselines covering 20--400\,m, and the receivers
tuned such that the upper sideband was centred on 870\,\um{}.  The
blazar, 1924$-$292, was used as the bandpass calibrator and Titan was
used for absolute flux calibration.  The nearby quasar, 1310+323, was
used to track the complex gains.

Following a tentative line detection using Zpectrometer on NRAO's
Green Bank Telescope, plausibly CO \J{1}{0} at \(z = 2.3078\)
\citep{2012ApJ...752..152H}, we obtained 4.4\,hr on-source using all
six of the 15-m antennas of the Institut de Radioastronomie
Millim\'{e}trique's Plateau de Bure Interferometer (IRAM's
PdBI).  These data were taken during 2011 January, in A
configuration, with the 2-mm receivers tuned to 139.380\,GHz,
corresponding to the rest-frame frequency of CO \J{4}{3} at that
redshift.  No CO line emission was visible within limits compatible
with the \J{1}{0} flux and further follow-up of HATLAS\,J132427 was
put on hold, pending a more robust redshift determination.  We use
these data here because they provide a usefully deep \(\left(
\sigma = 0.1 \, \mJy \, \text{beam}^{-1}\right)\),
high-resolution \(\left(0.68 \text{-arcsec} \times 0.45
\text{-arcsec, position angle (PA) 55}^{\circ}\right)\) 2-mm
continuum image.

\begin{figure}
  \centering
  \includegraphics[]{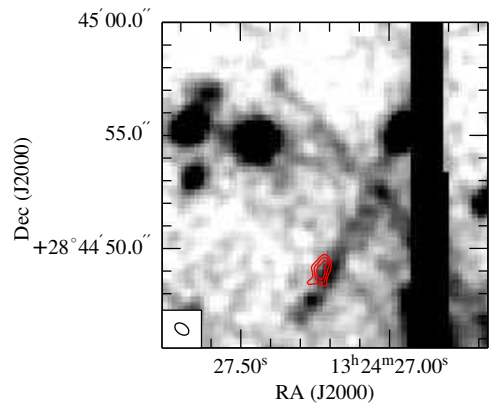}
  \caption{Continuum emission at 2\,mm, as observed by PdBI with a
    \(0.68\arcsec \times 0.45\arcsec \left(\text{PA 55}^{\circ}
    \right)\) synthesised beam, overlaid as contours on \(r\)-band
    imaging from the Canada-France-Hawaii Telescope
    \citep{2003ApJ...593...48G}.  The lensed arc is oriented NW--SE.
    The other elongated features are diffraction spikes caused by
    two bright, nearby stars. Both the brightest cluster galaxy and
    cluster centre are over 20\arcsec from the arc
    \citep{2005ApJS..157....1G}.}
\label{fig:optical}
\end{figure}

Via our interferometric imaging shown in Fig.~\ref{fig:optical} it
is immediately obvious that the far-IR source is coincident with the
$\sim 10$-arcsec giant arc reported by \citet{2003ApJ...593...48G},
strongly lensed by the rich foreground cluster RCS\,J132427+2845.2,
at \(z = 0.997 \pm 0.017\) \citep{2005ApJS..157....1G}.

\section{Far-IR and CO spectroscopy}
\label{spectroscopy}

\begin{table}
  \centering
  \caption{Continuum flux densities.}
  \begin{tabular}{rrr@{\extracolsep{0.167em}}lrl}
    \hline
    \hline
    \multicolumn{2}{c}
    {Wavelength}       & \multicolumn{3}{c}{\(S_\nu\)}  & Notes\\
    \hline
    100  & \um & 41   & $\pm \, 4^\text{a}$ & mJy  & \herschel{} PACS\\
    160  & \um & 180  & $\pm \, 14^\text{a}$ & mJy  & \herschel{} PACS\\
    250  & \um & 347  & $\pm \, 25^\text{b}$ & mJy  & \herschel{} SPIRE\\
    350  & \um & 378  & $\pm \, 28^\text{b}$ & mJy  & \herschel{} SPIRE\\
    500  & \um & 268  & $\pm \, 21^\text{b}$ & mJy  & \herschel{} SPIRE\\
    870  & \um & 30.2 & $\pm \, 5.2$         & mJy  & SMA\\
    2    & mm  & 1.2  & $\pm \, 0.1$         & mJy  & IRAM PdBI\\
    3.5  & mm  & 200  & $\pm \, 170$         & \uJy & CARMA\\
    4.3  & cm  & 350  & $\pm \, 30$          & \uJy & VLA\\
    21   & cm  & 1.95 & $\pm \, 0.24$        & mJy  & VLA (FIRST)\\
    \hline
  \end{tabular}
  \begin{tabular}{l@{\extracolsep{0.3em}}p{23em}}
    \(^\text{a}\) & Errors include 3- and 5-per-cent calibration
    uncertainties at 100- and 160-\um, respectively.\\
    \(^\text{b}\) & Errors include the contribution due to confusion and a
    7-per-cent calibration uncertainty has been added in
    quadrature (Valiante et al., in preparation).
  \end{tabular}
  \label{tab:fluxes}
\end{table}

\begin{figure}
  \centering
  \includegraphics[]{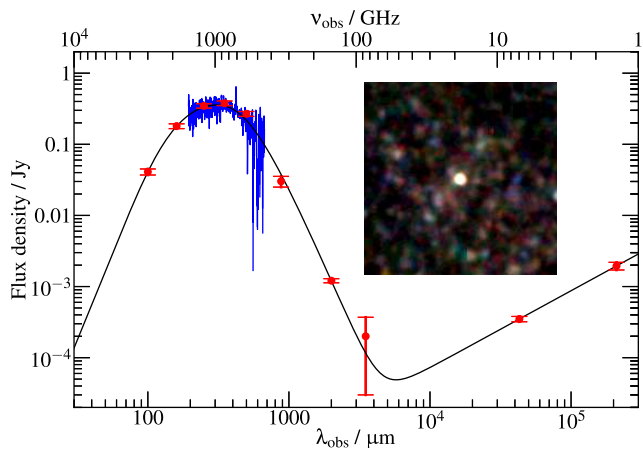}
  \caption{Far-IR-through-radio SED of HATLAS\,J132427, with the
    best-fit synchrotron and power-law temperature distribution dust
    model shown in black.  The FTS spectrum is shown in blue.
    \emph{Inset}: 15-arcmin $\times$ 15-arcmin false-colour image of
    HATLAS\,J132427 in the three \herschel{} SPIRE filters.}
  \label{fig:sed}
\end{figure}

\subsection{A blind redshift from the SPIRE FTS}
\label{fts}

During 2012 August 02, HATLAS\,J132427 was observed with a single
pointing of the SPIRE Fourier Transform Spectrometer (FTS) for
3.8\,hr, the spectrum covering \(\lambda_\text{obs} =
\text{194--671\,\um}\).  The data were processed using the \herschel\
data processing pipeline \citep{2010SPIE.7731E..99F} within the
\emph{Herschel Interactive Processing Environment} v10.  Spectra from
detectors arranged around the central `on-source' detector were
averaged to produce a local background measurement, and subtracted
from the source spectrum.

The \(^2\text{P}_{3/2} \rightarrow {}^2\text{P}_{1/2}\) \cii{}
transition at rest-frame 157.74\,\um{} is one of the brightest lines
seen in the far-IR waveband, providing up to 1 per cent of \lir{}
(measured across rest-frame 8--1{,}000\,\um) in star-forming galaxies
\citep{1991ApJ...373..423S}.  Within the spectral range covered by the
SPIRE FTS in our observations of lensed starburst galaxies, this line
is likely to be the most significant; indeed, it is often the only
transition detected \citep[e.g.][]{2010A&A...518L..35I}.

\begin{figure}
\centering
\includegraphics[]{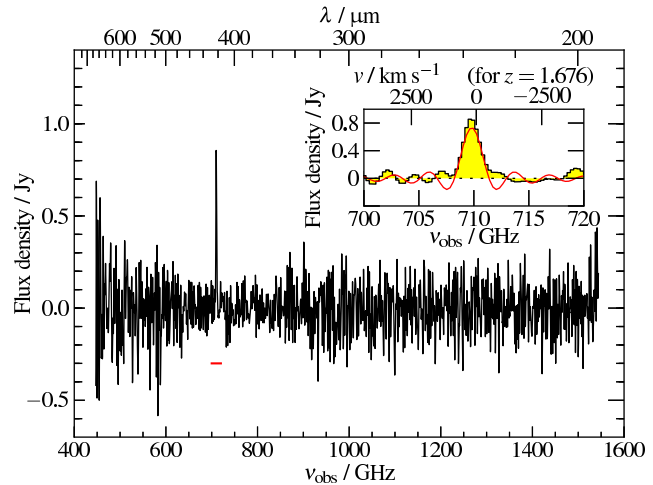}
\caption{The continuum-subtracted \herschel{} SPIRE FTS spectrum.  No
  other features correspond to expected transition lines.
  \emph{Inset}: Zoomed in on the region indicated by the red line in
  the parent plot.  Velocity scale corresponds to [C\,\textsc{ii}] at
  \(z = 1.676\).  The best-fit sinc profile is overlaid in red.}
\label{fig:fts_spec}
\end{figure}

The SPIRE FTS spectrum of HATLAS\,J132427, shown in
Fig.~\ref{fig:fts_spec}, displays a 7.5\,\(\sigma\) marginally
resolved (1.2\,GHz spectral resolution) emission line at
\((709.9 \pm 0.4) \, \text{GHz}\).  Attributing this to \cii{}
indicates a redshift, \(z=1.677 \pm 0.001\).  No other lines were
detected, with \(3 \sigma < 200\), 230, 150, 600 Jy \kms{} for
[O\,\textsc{iii}] 88\,\um{}, [N\,\textsc{ii}] 122\,\um{},
[O\,\textsc{i}] 145\,\um{}, and [N\,\textsc{ii}] 205\,\um{}
respectively, meaning that this redshift remained tentative at this
point.

\subsection{Redshift confirmation via CO}
\label{coobs}

\begin{figure}
\centering
\includegraphics[]{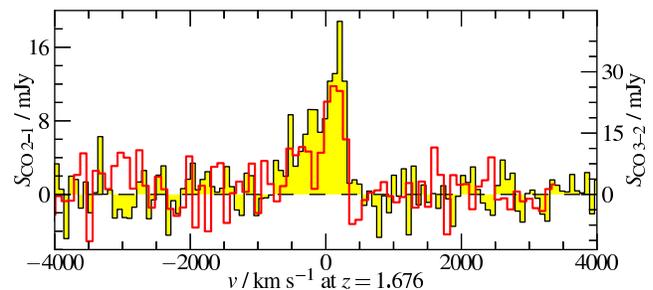}
\caption{CO \J{2}{1} spectrum from CARMA, binned to 20.8\,MHz, with
  the CO \J{3}{2} IRAM PdBI spectrum shown in red, binned to 40\,MHz
  and put on the same brightness temperature (\Tb) scale.}
\label{fig:co}
\end{figure}

To verify the SPIRE FTS redshift determination, we used the Combined
Array for Research in Millimeter-wave Astronomy (CARMA) to search for
the CO \J{2}{1} line (\(\nu_\text{rest}= 230.538 \, \text{GHz}\)),
which should be redshifted to approximately \(\nu_\text{obs} = 86.0 \,
\text{GHz}\) for \(z = 1.68\).  Observations were carried out using
the 3-mm receivers during 2012 November 23 in D configuration
(11--146\,m baselines), with 2.3\,hr spent on source.  The blazars,
1310+323 and 0927+390, were used for complex gain calibration and to
derive the bandpass shape.  Absolute flux densities should be accurate
to within \(\sim 15\) per cent.  We obtained an 8\,\(\sigma\)
detection of line emission, close to the expected frequency -- see
 Fig.~\ref{fig:co} -- thus confirming the redshift, with \(z = 1.676
\pm 0.001\).  The 34.8\,GHz GBT line is therefore presumably spurious.

We also imaged the source in CO \J{3}{2} with the IRAM PdBI, also
during 2012 November. We obtained 1.1\,hr of integration time, using
all six of the 15-m antennas, this time in D configuration -- the most
compact.  The observing frequency was set to 129.028\,GHz,
corresponding to the redshifted frequency of CO \J{3}{2}
(\(\nu_\text{rest} = 345.796 \, \text{GHz}\)) for
\(z = 1.68\).  Again, we found a bright 3\,\(\sigma\) emission line
at the expected frequency, (Fig.~\ref{fig:co}).

The two line profiles are consistent with one another; neither can be
described well with a single Gaussian, suggesting that their shape is
due to either a merger or a rotating, gas-rich disk
\citep[e.g.][]{2010ApJ...724..233E, 2013ApJ...772..137I}.  The line
width (deduced from fits using a single Gaussian), \((500 \pm 140) \,
\kms\) {\sc fwhm}, is typical of those seen for DSFGs
\citep{2005MNRAS.359.1165G}.

\section{Discussion}
\label{discussion}

\begin{table}
  \centering
  \caption{Properties of HATLAS\,J132427.}
  \begin{tabular}{l @{\hspace{4.5em}} r @{$\, \pm \, $} l}
    \hline
    \hline
    Property                     & \multicolumn{2}{c}{\hspace{-0.4em}Value}\\
    \hline
    $\Td\ /\K$                                      & 33.1 & 3.2\\
    $M_\text{d} \, / \, 10^9 \, \mu^{-1} \, \msun$  & 1.07 & 0.15\\
    ${A_\text{projected}}^\text{a}\,/\text{kpc}^2$  & 21.4 & 5.7\\
    $\lir \, /10^{13} \, \mu^{-1} \, \lsun$         & 2.91 & 0.12\\
    Synchrotron spectral index $\alpha$             & 1.08 & 0.09\\
    $q_\text{IR}$                                   & 1.85 & 0.09\\
    SFR$^\text{b} \,
        /\mu^{-1} \, \msun \, \text{yr}^{-1}$   & \multicolumn{1}{r}{4{,}300}
                                                    \hspace{0.42em}\\
    \cii{} $I_\cii \, /\text{Jy} \, \kms$           &  486 & 52\\
    \cii{} FWHM $/ \kms$                            &  800 & 64\\
    \cii{} $L_\cii \, /10^{10}\,\mu^{-1}\,\lsun$    & 5.80 & 0.62\\
    CO \J{2}{1} $I_\text{CO} \, /\text{Jy}\,\kms$   & 11.3 & 1.4\\
    CO \J{2}{1} FWHM $/ \kms$                       &  640 & 270\\
    CO \J{2}{1} $L_\text{CO}
      \, /10^8 \, \mu^{-1} \, \lsun$                & 1.64 & 0.20\\
    CO \J{3}{2} $I_\text{CO} \, /\text{Jy}\,\kms$   & 11.5 & 3.5\\
    CO \J{3}{2} FWHM$^\text{c} \, / \kms$           &  450 & 170\\
    CO \J{3}{2} $L_\text{CO}
      \, /10^8 \, \mu^{-1} \, \lsun$                & 2.50 & 0.76\\
    \R{3}{2}{2}{1}                                  & 0.45 & 0.15\\
    ${M_\text{gas}}^\text{d}\,
      /10^{11} \, \mu^{-1} \,\msun$                 &  3.6 & 0.6\\
    SFE$^\text{d} \, /\lsun \, \msun^{-1}$          & \multicolumn{1}{r}{83}
                                                      \hspace{0.42em}\\
    \hline
  \end{tabular}
  \begin{tabular}{l@{\extracolsep{0.3em}}p{24em}}
    \(^\text{a}\) & From fitting SED to dust emission.\\
    \(^\text{b}\) & Following \citet{2011ApJ...737...67M}.\\
    \(^\text{c}\) & From fitting a Gaussian with its mean constrained to
                    that of the \J{2}{1} fit.\\
    \(^\text{d}\) & Average determined via CO \J{2}{1} and \J{3}{2}.
  \end{tabular}
  \label{tab:properties}
\end{table}

Characterisation of the dust emission of HATLAS\,J132427 was performed
by fitting the power-law dust temperature model of
\citet{2010ApJ...717...29K} to the continuum flux densities detailed
in Table~\ref{tab:fluxes}.  Derived quantities are detailed in
Table~\ref{tab:properties}.

Of the \(350 \pm 30 \, \uJy\) integrated flux density at 7\,GHz,
\(270 \pm 25 \, \uJy\) ($77\pm 10$ per cent) is in a compact
component at RA = 13:24:27.225, Dec.\ = +28:44:49.01 (\(\pm 0.02\),
J2000), with a best-fit deconvolved size of \(0.45 \pm
0.05\)\,arcsec \(\times \; 0.09 \pm 0.05\)\,arcsec, PA \(152 \pm
4^{\circ}\). Comparing this with the 2-mm emission (RA = 13:24:27.229
\(\pm\) 0.004, Dec.\ = +28:44:49.07 \(\pm\) 0.06, J2000, \(1.0 \pm
0.2\)\,arcsec \(\times <0.3\)\,arcsec, PA \(152 \pm 9^{\circ}\))
suggests that the radio emission is significantly more compact than
the dusty star-forming material, suggestive of a non-starburst origin.
The radio spectral index, $\alpha$ (where $S_\nu\propto\nu^\alpha$) is
steeper than the typical $\alpha=-0.8$, adding credibility to the idea
that an AGN dominates the radio emission
\citep[e.g.][]{2010MNRAS.401L..53I}.  Adopting the widely-used
parametrisation of the far-IR/radio correlation, \(q_\text{IR}\)
(with \lir{} measured across rest-frame 8 -- 1{,}000 \um) we find a
value ($1.85 \pm 0.09$) rather lower than that usually seen for DSFGs,
\(\approx 2.4\) \citep{2001ApJ...554..803Y, 2010A&A...518L..31I,
2010A&A...518L..35I}, again consistent with an AGN-related
contribution to the radio luminosity
\citep[e.g.][]{2005ApJ...634..169D}.

The [O\,\textsc{iii}] 88\,\um{} line was not detected.  An
[O\,\textsc{iii}] line of similar SNR to \cii{} was observable from
SDP81 \citep[which also has \(q_\text{IR} \sim 1.8\pm 0.1\)
--][]{2011MNRAS.415.3473V}, this time suggesting that the AGN has a
greater influence in SDP81 than in HATLAS\,J132427
\citep{2009ApJ...701.1147A}.

CO is commonly used to trace molecular gas reservoirs.  Estimates
of the \J{1}{0} line flux can be derived from the higher-\(J\) line
fluxes assuming Rayleigh-Jeans brightness temperature (\Tb) ratios,
e.g.\ \(\R{2}{1}{1}{0} \sim 0.84 \pm 0.13\) and \(\R{3}{2}{1}{0} \sim
0.52 \pm 0.09\) \citep{2011MNRAS.412.1913I, 2013MNRAS.429.3047B}:
consistent with the \Tb\ ratio measured here for
\(\R{3}{2}{2}{1}\), \(0.45 \pm 0.15\).  Adopting \(\alpha_\text{CO} =
0.8 \, \xunits\), as is usual for starburst galaxies
\citep[e.g.][]{2013arXiv1301.3498B}, we derive a \(\text{H}_2 +
\text{He}\) mass, \(M_\text{gas} = (3.6 \pm 0.6) \times 10^{11} \,
\mu^{-1} \, \msun\), having averaged the \J{2}{1} and
\(3 \! \rightarrow \! 2\) measurements.  Following
\citet{2012ApJ...752..152H} and \citet{2013MNRAS.429.3047B}, a rough
magnification estimate can be derived from the \J{1}{0} luminosity and
linewidth, yielding \(\mu \approx 11\).

\begin{figure}
\centering
\includegraphics[]{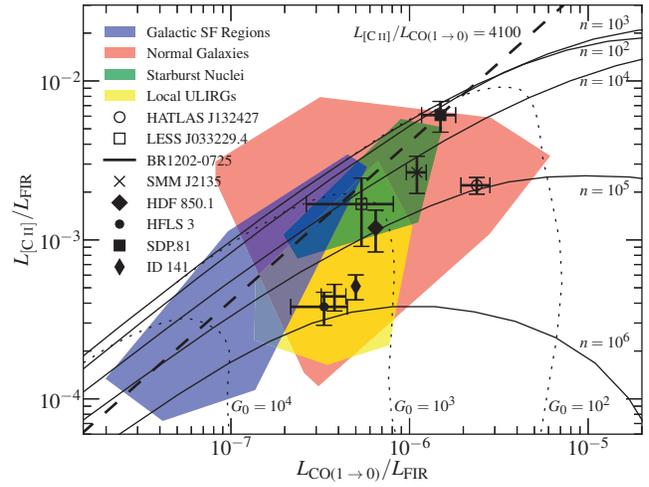}
\caption{Diagnostic plot, adapted from \citet{2010ApJ...714L.162H},
with identical corrections applied to \(L_\cii\) (\(\times 0.7\) --
non-PDR emission) and \(L_{\text{CO1-0}}\) (\(\times 2.0\) -- an
optically thick transition).  Shaded regions are approximate coverage
of local objects.  Neglecting differential lensing magnification,
\(n\) and \(G_0\) estimates are magnification-independent.  Also
shown are other high-redshift starbursts:
LESS\,J033229.4$-$275619 \citep{2011A&A...530L...8D},
BR1202$-$0725 \citep{2012ApJ...752L..30W},
SMM\,J2135$-$0102 \citep{2010A&A...518L..35I}, HDF\,850.1
\citep{2012Natur.486..233W}, HFLS\,3 \citep{2013Natur.496..329R},
\emph{H}-ATLAS SDP.81 \citep{2011MNRAS.415.3473V}, and HATLAS ID\,141
\citep{2011ApJ...740...63C}.  Where necessary, literature values have
been converted using \(\R{2}{1}{1}{0} \sim 0.84 \pm 0.13\), and the
ratios obtained by integrating under the HATLAS\,J132427 SED:
\(L_{\text{FIR} (42.5  \text{--} 122.5 \um)} =
0.69 \times L_{(40 \text{--} 500 \um)} =
0.63 \times L_{\text{IR} (8 \text{--} 1000 \um)} \).  }
\label{fig:diagnostic}
\end{figure}

An independent estimate of the gas mass can be derived via the
gas-to-dust mass ratio, \(M_\text{gas} / M_\text{d}\).  For 1/3 Solar
metallicity, this is \(220^{+180} _{-90}\)
\citep{2012arXiv1212.1208S}, suggesting \(M_\text{gas} = 2.4^{+2.0}
_{-1.0} \times 10^{11} \, \mu^{-1} \,\msun\), consistent with our
CO-based estimate, albeit with a large uncertainty.  These estimates
of the gas mass correspond to a star-formation efficiency
\(60 \lesssim \lir /M_\text{gas} \lesssim 220 \, \lsun \,
\msun^{-1}\), and a gas-consumption timescale \(M_\text{gas} /
\text{SFR} \sim 30 \text{--} 100 \, \text{Myr}\), consistent with
those typically measured for DSFGs \citep[e.g.][]{2005MNRAS.359.1165G}.

Using the PDR model displayed in Fig.~\ref{fig:diagnostic},
representing the far-UV illuminated surface of an interstellar cloud,
we estimate \(n \sim 10^{4.8} \, \text{cm}^{-3}\) and
\(G_0 \sim 10^{2.3}\), a density and far-UV radiation field
strength comparable with other high-redshift starbursts.
These models assume all CO emission is produced in PDRs, neglecting
any contribution from quiescent gas or turbulent heating which provide
plausible explanations for the comparatively large \(L_\text{CO(\(1\!
\rightarrow \! 0\))} / L_\text{FIR}\) in this object, as may
increased metallicity.  Differential magnification might also affect
these derived properties, if the CO and far-IR emission are not
co-spatial \citep[e.g.][]{2012ApJ...753..134F}.

Without a detailed mass model of the cluster lens, it is not possible
to investigate the extent of differential magnification of the DSFG,
but we can nonetheless attempt a plausibility test of the statistical
likelihood.  Following \citet{2012MNRAS.424.2429S}, we modelled a
singular isothermal ellipsoid with a 20 arcsecond critical radius and
ellipticity of 0.2, with the same submm galaxy model used in that
paper.  As with the galaxy-scale lenses in
\citet{2012MNRAS.424.2429S}, we found Fig.~\ref{fig:diagnostic} to be
largely unaffected by differential magnification, even
for \(\mu > 10\).  However, the aspect ratio of the observed-frame
optical emission suggests a high magnification and large angular
extent, implying long-baseline interferometric measurements may
resolve out some of the emission, as discussed in \S\ref{coobs}.

\section{Implications for far-IR integral-field spectroscopy}
\label{implications}

The Space Infrared Telescope for Cosmology and Astrophysics
\citep[SPICA --][]{2009ExA....23..193S} is a planned successor to
\herschel.  Thermal emission from the 80\,\K{} primary telescope of
\herschel{} is six orders of magnitude brighter than the far-IR
background; with a dish temperature of 6\,\K, SPICA should be two
orders of magnitude more sensitive than any previous far-IR facility.

SAFARI -- the SpicA FAR-infrared Instrument -- is an imaging FTS
proposed for SPICA, covering 34--210\,\um{} \(\left([\text{O}
\,\textsc{i}] \; 63 \, \um \; \text{to} \; z = 2.3\right)\) over a
\(2\arcmin \times 2\arcmin\) field of view with an intended spectral
resolution \(\text{R} \sim 2000\) and sensitivity \(< \text{few}
 \times 10^{-19} \, \text{W} \, \text{m}^{-2}\)
\citep[\(5\sigma: 1 \,\text{hr}\);][] {2009ExA....23..193S}.

Our observation of \cii\ from HATLAS\,J132427 provides \(I_{\cii} =
(1.15 \pm 0.12) \times 10^{-17} \, \text{W} \, \text{m}^{-2}\).  An
[O\,\textsc{i}] line of this flux will likely be observable to several
\(\sigma\) within two seconds of integration with SPICA.  While such
sources will likely be known before a SPICA pointing, the ability of
\herschel{} to obtain a blind redshift provides confidence that with
observations of order 1\,hr, the much more sensitive SPICA will
pinpoint redshifts of multiple unlensed few-mJy CIB sources
\citep[potentially several from each spectrum --][]
{2007A&A...465..125C, 2010PASJ...62..697R}, reducing the
need for the blind CO line searches typically required today.  With
such integration times, blind fine-structure line detections are
feasible and a gateway to detailed imaging/dynamical studies with ALMA.

\section{Conclusions}
\label{conclusions}

We present the first blind redshift to have been obtained via
\herschel{} spectroscopy.  The galaxy, at \(z = 1.68\), is lensed by a
rich galaxy cluster at \(z = 1.0\) into an arc of length \(\sim 10
\text{-arcsec}\).  A lensing model will be presented in Fu et al.{}
(in preparation).

This source appears to be a disk-like system, with a best-fit \lir{}
of \(2.91 \times 10^{13} \, \mu^{-1} \, \lsun\), dust temperature of
\(\Td = 33 \K\), and SFR of \(4{,}300 \, \mu^{-1} \, \msun\,
\text{yr}^{-1}\), and a compact AGN-related contribution to the radio
flux density.

The observed \cii{} flux is \((1.15 \pm 0.12) \times 10^{-17} \,
\text{W} \, \text{m}^{-2}\), suggesting an [O\,\textsc{i}] line of
similar flux could be detected by SPICA to \(5\sigma\) with an
integration time of order 3\,sec.  Even many unlensed DSFGs may
require no more than $\approx 1$\,hr of integration.  The feasibility
of blind line detections will not only directly advance our knowledge
of high-redshift systems and their interstellar media, but act as a
gateway to detailed imaging and spectral studies with
\mbox{(sub-)}millimetre interferometers such as ALMA.

\section*{Acknowledgements}

We thank Howard Yee and Michael Gladders for providing the CFHT image.
US participants in \emph{H}-ATLAS acknowledge support from NASA
through a contract from JPL.  RJI acknowledges support from the
European Research Council (ERC) in the form of Advanced Grant,
\textsc{cosmicism}.  \herschel{} was an ESA space observatory with
science instruments provided by European-led Principal Investigator
consortia and with important participation from NASA.   IRAM is
supported by INSU/CNRS (France), MPG (Germany) and IGN
(Spain).  Support for CARMA construction was derived from the states
of California, Illinois, and Maryland, the James S. McDonnell
Foundation, the Gordon and Betty Moore Foundation, the Kenneth T. and
Eileen L. Norris Foundation, the University of Chicago, the Associates
of the California Institute of Technology, and the National Science
Foundation.  Ongoing CARMA development and operations are supported by
the National Science Foundation under a cooperative agreement, and by
the CARMA partner universities.  The NRAO is a facility of the NSF
operated under cooperative agreement by Associated Universities,
Inc.  The SMA is a joint project between the Smithsonian Astrophysical
Observatory and the Academia Sinica Institute of Astronomy and
Astrophysics and is funded by the Smithsonian Institution and the
Academia Sinica.

\bibliography{blind_redshift}

\bsp 
\end{document}